**Title:** Localized FDG loss in lung cancer lesions

**Authors:** Davide Parodi, Edoardo Dighero, Giorgia Biddau, Francesca D'Amico, Matteo Bauckneht, Cecilia Marini, Sara Garbarino, Cristina Campi, Michele Piana, Gianmario Sambuceti

**Authors information**

Davide Parodi: Università Campus Bio-Medico di Roma and Dipartimento di Matematica Università degli Studi di Genova, Genova (Italy)

Edoardo Dighero: Dipartimento di Scienza della Salute, Università degli Studi di Genova, Genova (Italy)

Giorgia Biddau: Dipartimento di Matematica, Università degli Studi di Genova, Genova (Italy)

Francesca D'Amico: Dipartimento di Scienza della Salute, Università degli Studi di Genova, Genova (Italy)

Matteo Bauckneht: Dipartimento di Scienza della Salute, Università degli Studi di Genova; Nuclear Medicine Unit, IRCCS Ospedale Policlinico San Martino, Genova (Italy).

Cecilia Marini: Institute of Molecular Bioimaging and Physiology, National Research Council (CNR); Nuclear Medicine Unit, IRCCS Ospedale Policlinico San Martino, Genova (Italy).

Cristina Campi: Dipartimento di Matematica, Università degli Studi di Genova; Life Science Computational laboratory, IRCCS Ospedale Policlinico San Martino, Genova (Italy).

Sara Garbarino: Life Science Computational laboratory, IRCCS Ospedale Policlinico San Martino, Genova (Italy).

Michele Piana: Dipartimento di Matematica, Università degli Studi di Genova; Life Science Computational laboratory, IRCCS Ospedale Policlinico San Martino, Genova (Italy).

Gianmario Sambuceti, Dipartimento di Scienza della Salute, Università degli Studi di Genova; Nuclear Medicine Unit, IRCCS Ospedale Policlinico San Martino, Genova (Italy).

**Corresponding author:** Sara Garbarino, sara.garbarino@hsanmartino.it

**Acknowledgments:** MP acknowledges the support of the PRIN PNRR 2022 Project 'Inverse Problems in the Imaging Sciences (IPIS)' 2022ANC8HL, cup: D53D23005740006

# Abstract

**Purpose**: Analysis of [18F]-Fluorodeoxyglucose (FDG) kinetics in cancer has been most often limited to the evaluation of the average uptake over relatively large volumes. Nevertheless, tumor lesion almost contains inflammatory infiltrates whose cells are characterized by a significant radioactivity washout due to the hydrolysis of FDG-6P catalyzed by glucose-6P phosphatase. The present study aimed to verify whether voxel-wise compartmental analysis of dynamic imaging can identify tumor regions characterized by tracer washout.

**Materials & Methods**: The study included 11 patients with lung cancer submitted to PET/CT imaging for staging purposes. Tumor was defined by drawing a volume of interest loosely surrounding the lesion and considering all inside voxels with standardized uptake value (SUV) >40% of the maximum. After 20 minutes dynamic imaging centered on the heart, eight whole body scans were repeated. Six parametric maps were progressively generated by computing six regression lines that considered all eight frames, the last seven ones, and so on, up to the last three.

**Results**: Progressively delaying the starting point of regression line computation identified a progressive increase in the prevalence of voxels with a negative slope. In the most delayed parametric map, these voxels accounted for 0.5%-4.5% (median value 2%) of tumor volume. This behavior was independent of tumor size and was most preferentially located at the lesion borders with respect to its core.

**Conclusions**: The voxel-wise parametric maps provided by compartmental analysis permits to identify a measurable volume characterized by radioactivity washout. The spatial localization of this pattern is compatible with the recognized preferential site of inflammatory infiltrates populating the tumor stroma and might improve the power of FDG imaging in monitoring the effectiveness of treatments aimed to empower the host immune response against the cancer.

# Introduction

The widespread use of 18F-fluoro-deoxyglucose (FDG) in Positron Emission Tomography (PET) imaging in cancer patients largely relies on its user-friendly procedure that allows mapping tracer uptake in the whole body with a single image acquisition in the late steady state phase. Indeed, FDG shares with glucose both GLUT-facilitated transmembrane transport and hexokinase-catalyzed phosphorylation to FDG-6P that, in turn, irreversibly accumulates within the cytosol being a false-substrate for the enzymes channeling glucose-6P to glycolysis or pentose phosphate pathway [1]. In agreement with this kinetic model, current guidelines strictly recommend not to shorten the time between tracer injection and scanning below 55 minutes [2-4]. By contrast, a larger variability is allowed for the maximal duration of this interval, that can be extended by a further 30 minutes beyond the optimal time of one hour [5].

This kinetic model implicitly assumes that FDG-vehiculated radioactivity cannot be lost by cancer lesions. At the time of the introduction of FDG-PET modality, this concept fitted the notion that glucose-6P-phosphatase (G6Pase) – the enzyme hydrolyzing the sequestered glucose-6P and FDG-6P to the freely exchangeable glucose and FDG – is only expressed in liver, gut, and kidneys [6,7]. However, more recent literature on glycogen storage diseases discovered the G6Pase isoform β (or G6PC3) that is ubiquitously expressed in all tissues [8] and in several tumors, including uterus, lung, breast, and colon cancer [9, 10, 11] as well as glioblastoma [12]. The activity of this enzyme and its capability to regulate the glucose-6P/glucose ratio has been found to be extremely relevant in the maturation and proliferation of mesenchymal cells [13] and in the modulation of the host inflammatory reaction [14-16].

A large literature now documents that virtually all solid tumors are infiltrated by lymphocytes and macrophages [17]. The high G6Pase expression of these cells and their variegate distribution might thus configure tracer accumulation as a reversible process in some tissue volumes.

So far, this hypothesis has never been tested since FDG accumulation rate in cancer has been most often evaluated by analyzing the time-concentration curves of relatively large volumes of interest (VOIs) whose extension did not permit the identification of any possible heterogeneity in tracer kinetics [18]. With the advancement of PET/CT technology it is now possible to simultaneously define tracer concentrations in both the arterial blood and in the tissues with serial whole-body acquisitions. By analyzing these time activity curves (TACs), we demonstrated that, contrary to commonly accepted models, lung cancer lesions exhibit measurable and heterogeneously distributed tracer loss.

# Methods

## Patient population and dynamic PET/CT acquisition

The study included 11 patients (9 men and 2 women, mean age 65 years, median 68 years, range 29 – 88 years) submitted to whole-body PET examination for staging of suspected lung cancer. Imaging was performed in the early morning, after 12 hours of fasting. After measurement of body weight and serum glucose level, an antecubital vein was cannulated, and each patient was positioned on the bed of a Siemens Flow mCT40 system (Siemens, Erlangen, Germany) to undergo the preliminary X-ray CT scanning performed according to the conventional procedure [2]. A list-mode acquisition was started soon before the bolus injection of FDG (0.8 MBq/Kg body weight) with the field of view focused on the heart for 20 minutes. Immediately thereafter, acquisition mode was shifted, and seven whole body passages were performed from the skull to the mid-tights at the speed of 1.4 mm/sec. A last equilibrium scan completed the acquisition procedure at the standard bed speed of 1 mm/sec to be analyzed for the clinical report.

## Image Analysis

The chest-centered part of the dynamic acquisition was binned according to the following frame sequence: 12 x 5 secs, 12 x 10 secs, 8 x 15 secs, 6 x 30 secs, 2 x 60 secs, 5 x 120 secs. By contrast, the eight subsequent whole-body scans were characterized by a slightly variable time sequence due to the length of the desired field of view. Accordingly, the acquisition time of each slice was defined based on the DICOM metadata to perform an accurate correction of $^{18}$F physical decay.

An expert nuclear physician thus identified a ≥5 mL VOI on the descending aorta to estimate the input function (IF) defined by the activity concentration in the arterial blood at all times of the dynamic chest-centered frames and in the subsequent whole-body acquisitions.

A further VOI was drawn to loosely surround the tumor lesion in the last scan, and a mask was created to set all the outside voxels to 0. Data were transformed into standardized uptake value (SUV) images according to the conventional formulation [19] and the cancer lesion was defined as the set of all voxels with radioactivity concentration >40% of the maximum value within the identified VOI.

## Parametric image analysis

For each patient, two sets of parametric images were set up. For the former, a Time Activity Curve (TAC) was generated in each tumor voxel throughout the eight whole-body acquisitions and six regression lines were computed considering all eight frames (1-8), the last seven ones (2-8), and so on, up to the last three (6-8). This analysis provided a first voxel-resolved description of FDG kinetics, made of a set of six parametric TAC images denoted, from now on, as $TAC_{1-8}$, $TAC_{2-8}$, ..., $TAC_{6-8}$. A regression analysis performed for each parametric TAC image allowed defining the involved voxels as accumulating when the slope of the regression line is positive and as releasing when the slope of the regression line is negative.

This preliminary evaluation was then compared with the conventional counterpart represented by the graphical approach to the compartmental analysis described by Patlak et al [20]. According to this largely adopted model, the tracer is freely exchanged between the blood and a series of intermediate reversible pools that act as an entry gate for the irreversible compartment in which entered radioactivity can never escape. Once the equilibrium between blood and reversible compartments is reached at time $t^*$, the irreversible accumulation within any given voxel is described by the equation:

$$C_T(t) = K_i \int_0^t C_b(\tau)d\tau + V_0 C_b(t) \qquad (1)$$

where $K_i$ represents the net accumulation rate in the irreversible compartment (clearance), $C_T(t)$ and $C_b(t)$ indicate the tracer concentrations in tissue and blood, respectively, while $V_0$ is the volume of the reversible compartments. Dividing both sides for $C_b(t)$, equation (1) can be re-written as:

$$\frac{C_T(t)}{C_b(t)} = K_i \frac{\int_0^t C_b(\tau)d\tau}{C_b(t)} + V_0 \, . \qquad (2)$$

A *standard Patlak analysis,* performed against all the dynamical FDG-PET images, allowed the computation of the *standard Patlak parametric image* whose voxels contain the corresponding $K_i$ values. However, reproducing the descriptive evaluation reported above, this same analysis was repeated six times, once again considering tumor radioactivity concentration at all six frames, in the last seven, up to the last three. Accordingly, in this *time-resolved Patlak analysis* six parametric images were set up, each one reporting the $K_i$ slope of the corresponding regression line, which was denoted as $(K_i)_{1-8}$ up to $(K_i)_{6-8}$. Once again, each voxel was identified as accumulating or releasing according to the sign of the corresponding regression slope.

## Statistical Analysis

All data are reported as mean ± standard error over mean (SEM). The statistical significance of the regression analysis in the case of the time-resolved approach was assessed by means of $R^2>0.3$ for both the TAC parametric images and the Patlak parametric images. In the correlation analysis $p<0.05$ was considered statistically significant.

# Results

## Clinical data

No patient showed metastatic lesions, while homolateral lymph nodes were involved in two patients. Tumor localization and final diagnosis at the histological examination of the harvested lesion is reported in Table 1. The same table also reports the volumes of both surrounding VOI and included tumor lesion as well as FDG uptake data. Overall, the final diagnosis was adenocarcinoma and squamous cell carcinoma in 7 and 4 patients, respectively. Maximal SUV was similar in the two histological types (14.5±10.1 vs 14.2±8.0, respectively, p=ns); by contrast, tumor volume was slightly, though not significantly lower in adenocarcinoma lesions with respect to squamous ones (20.7±17.7 mLs vs 71.3±75.9 mLs, respectively, p=.11)

## Standard and Time-resolved Patlak analyses confirm established results at volume level

Figure 1 illustrates the results of both the standard (1a) and the time-resolved (1b) Patlak analysis when the slope values of the regression lines are defined for the whole volume lesion. The $K_i$ value provided by the standard procedure nicely correlated with the average lesion SUV in all patients but one (#8) (1a). Similarly, the agreement between average $K_i$ values and average SUV remained invariant regardless the starting point of the time-resolved regression analysis (1b).

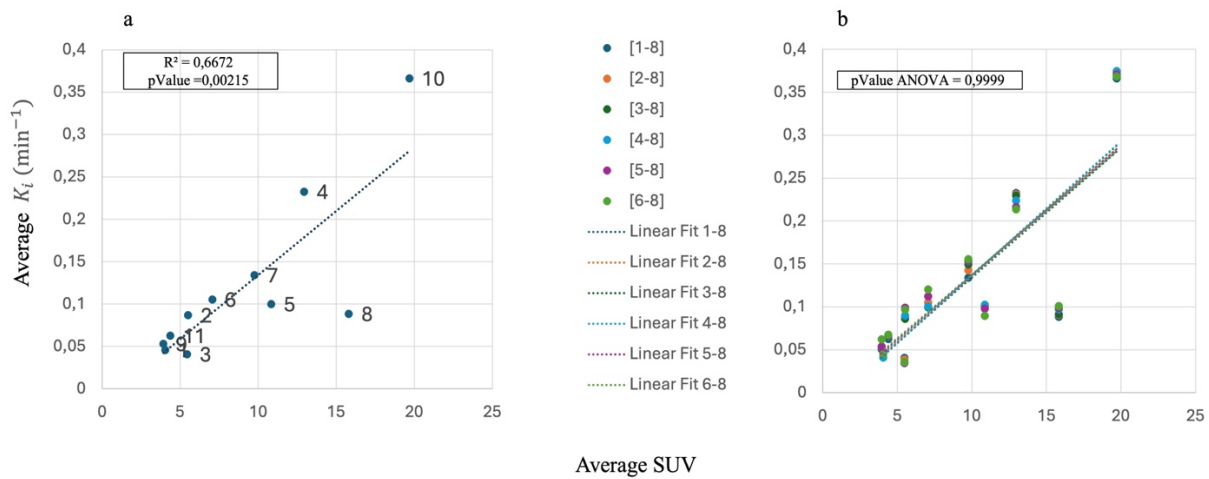

**Fig. 1** Correlation analysis between individual regional SUV and Patlak slope. (**a**) For each patient (labelled by their own ID), the x axis contains the SUVs averaged across the cancer volume computed at the last scan; the y axis contains the average values of the of standard Patlak slope. (**b**) For each patient, each colored dot represents the average value of the Patlak slope corresponding to a specific interval in the time-resolved analysis.

The result of this time-resolved analysis was further corroborated by the trend of average Ki values in the whole population (Fig. 2a) or in each subject (Fig. 2b) as a function of the starting point of regression analysis. This preliminary analysis confirmed the acknowledged model considering FDG uptake as an irreversible process when the whole lung cancer lesion is analyzed. However, this observation was not confirmed when the spatial resolution of the analysis was escalated to the voxel level (see next Sub-section).

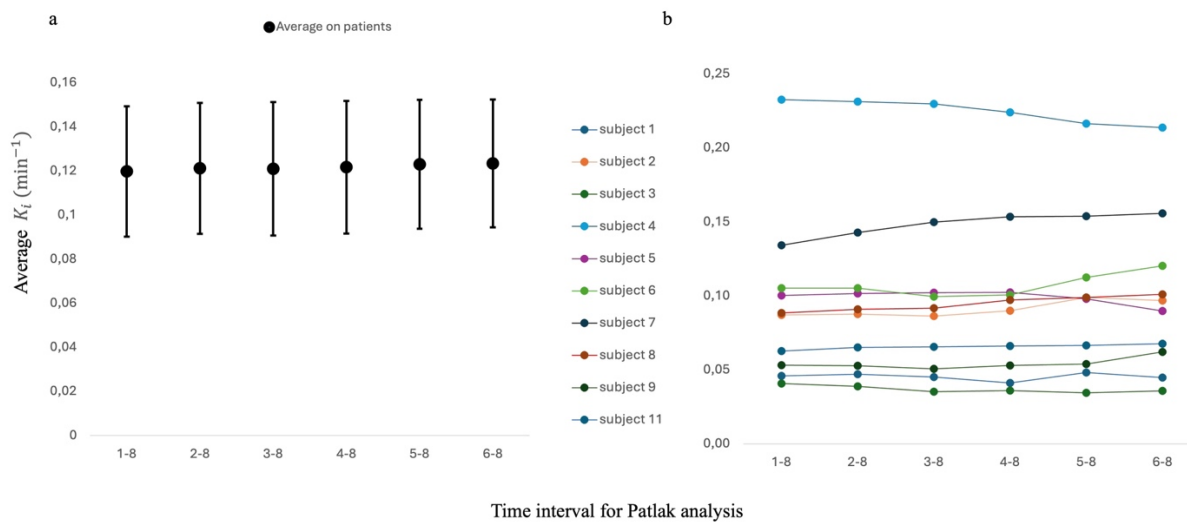

**Fig. 2** Volumetric time-resolved Patlak analysis. (**a**) For each time frame on the x axis the values on the y axis are the volumetric Patlak slopes (mean ± SEM) averaged across subjects. (**b**) The slopes on the y axis refer to each single subject

## Time-resolved Patlak analysis highlights releasing voxels

The analysis of parametric maps with a voxel-wise resolution provided a significantly different picture a shown in Figure 3. Indeed, the time-resolved Patlak analysis based on progressively delaying the starting point, documented the existence of a significant amount of releasing voxels, i.e., with consistently negative $K_i$ values. Specifically, the later the starting frame considered for the regression line computation, the higher the prevalence of cancer voxels characterized by a negative slope. Intriguingly, as documented by Figure 4, the slope of the identified regression line remained invariant in releasing voxels (Fig. 4a) and instead showed a progressive increase in the accumulating ones (Fig. 4b).

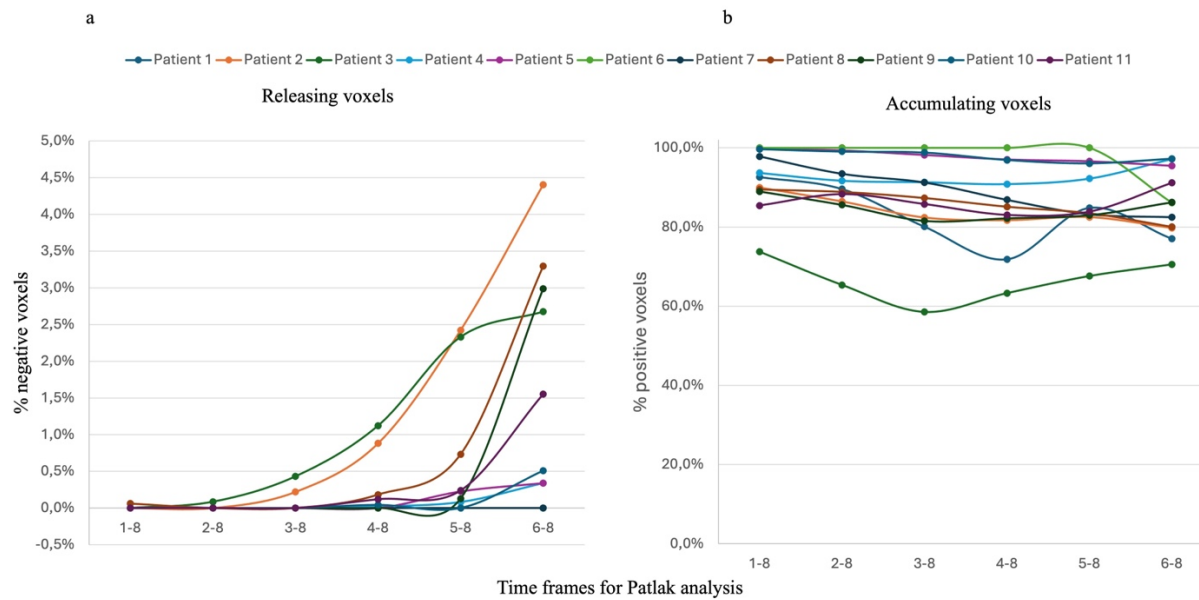

**Fig. 3** Time- and voxel-resolved Patlak analysis. The two plots show, for each patient and each time frame, the rate of voxels with negative (**a**) and positive (**b**) Patlak slope. Rates are computed with respect to the total number of voxels in the lesion

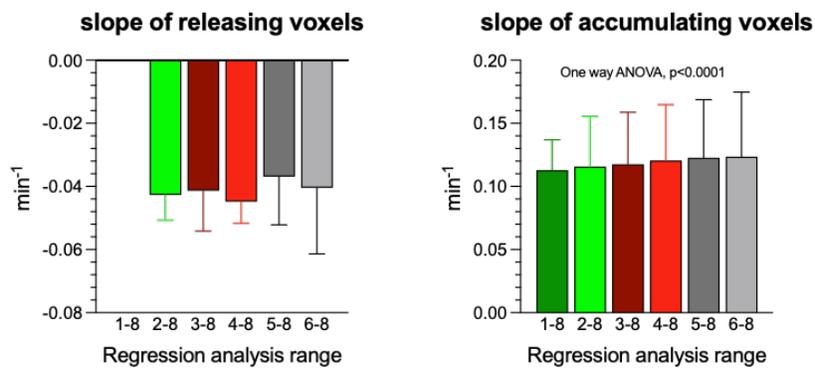

**Fig. 4** Distribution of regression line time- and voxel -resolved Patlak analysis slope in releasing (**a**) and accumulating voxels (**b**).

As shown in Figure 5, the progressive appearance of radioactivity release from a significant volume fraction was not related to an overestimation of tracer concentration in the arterial blood, since the same behavior was documented by the analysis of time-activity curves (TACs) in all voxels included in the cancer lesion. Finally, Figure 6 shows the correlations between the rate of releasing voxels and the cancer volume as computed using the static FDG-PET image.

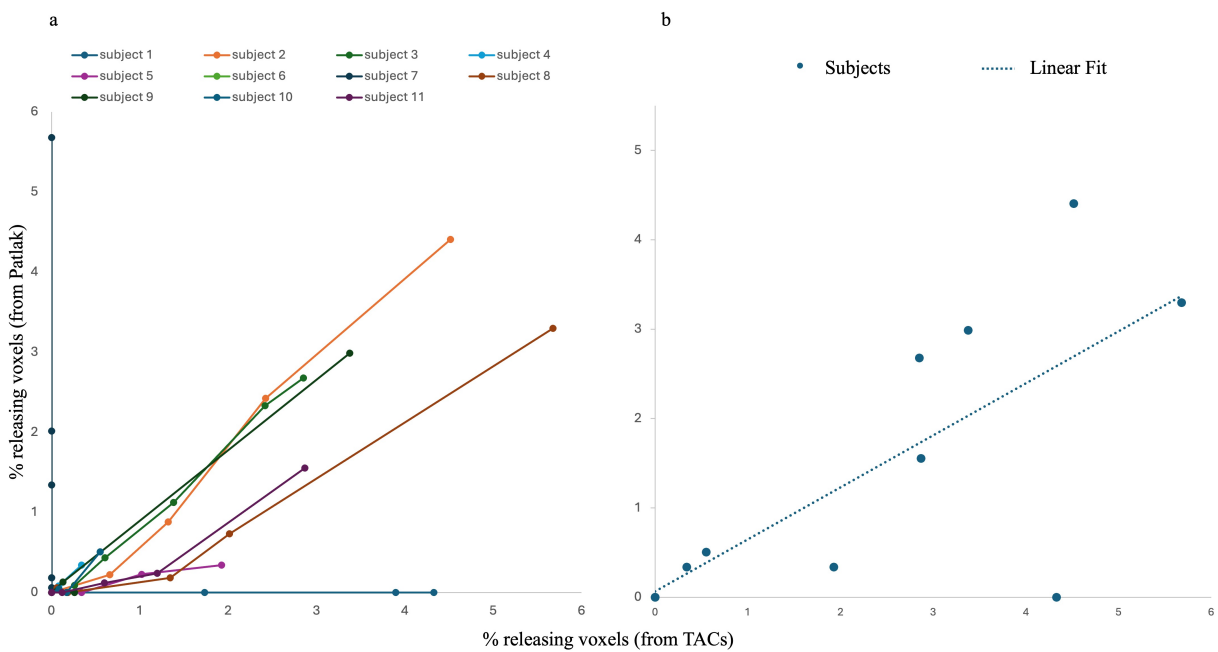

**Fig. 5** Comparison between the rate of releasing voxels computed by using the TAC slopes (x axis) and the Patlak slopes (y axis). (**a**) Each color corresponds to a subject and each point corresponds to a time frame. (**b**) Focus on frame 6-8, where each point corresponds to a patient.

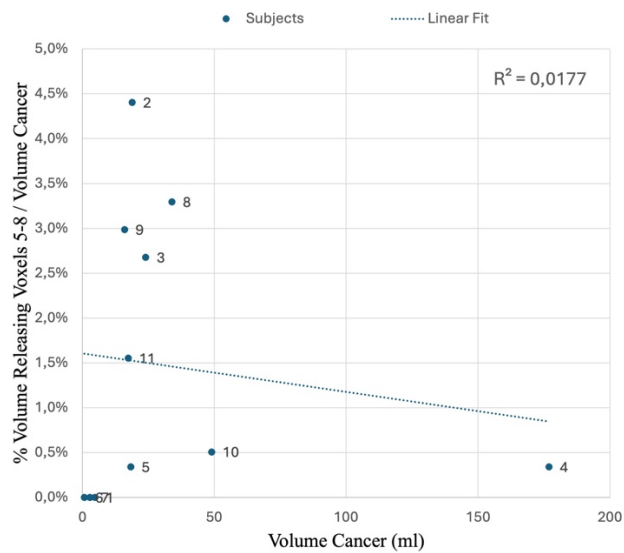

**Fig. 6** The rates of releasing voxels as computed from Patlak at interval 6-8, plotted against the cancer volume (in ml). Each patient is a dot marker. The dotted line is the linear trend on the subjects

This result shows that radioactivity release was not explained by a significant interference of normal tissues contaminating the analyzed voxels, since the extent of releasing regions was independent of tumor volume.

# Localization of the accumulating and releasing voxels

To document whether tracer kinetics was clustered in specific cancer regions, Figure 7 shows the voxels representing tumor boundaries, which were identified as all voxels with faces, edges, or corners adjacent to non-cancer ones (with SUVs <40% of maximum). By contrast, the core region was defined as the complementary tumor volume.

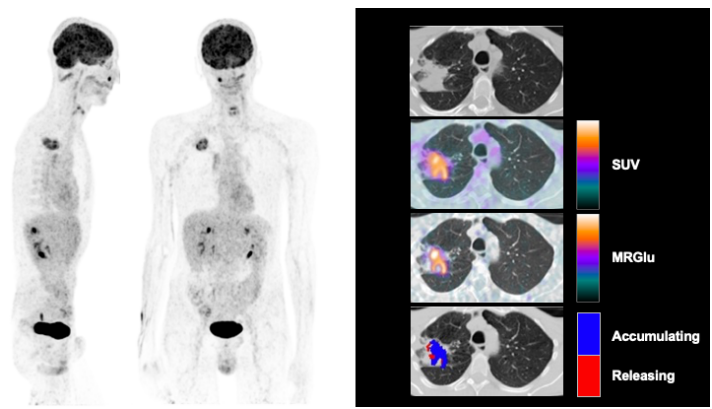

**Fig. 7** Spatial distribution of releasing and accumulating voxels co-registered with the CT scan of one of the analysed patients.

# Discussion

In the present study, the conventional Patlak analysis showed a heterogeneous FDG kinetics in treatment-naïve lung cancer. The tracer irreversibly accumulated in the largest part of the lesion volume; however, it was retained in a reversible pool for a measurable number of voxels. This behavior reflected a true tracer washout, as documented by the progressive decrease of the ratio between tumor and blood radioactivity concentration that ruled out a significant contamination by the blood volume included in the analyzed voxel [21]. It was independent of the lesion size and relatively more represented in the border zone with respect to the lesion core. Altogether, these findings indicate that the well documented heterogeneity of cell types populating the tumor eventually result in a heterogeneous kinetics of tracer accumulation as a possible index of a local infiltration by FDG-releasing inflammatory cells.

## Spatial resolution and FDG kinetics

As requested by the standard procedure, the Patlak regression line was computed setting the X values according to the full-time input function, i.e., from the radiotracer injection time until the dynamic scan is recorded, and plotting the Y values after the steady state has been reached. Due to the irreversible nature of FDG uptake, the estimated regression line in each voxel should have maintained constant values for both the slope and intercept, regardless the starting time of the analysis. This tenet was not respected in those lesion voxels that showed a measurable and progressive decrease in tumor/blood radioactivity ratio.

Due to its late occurrence and its relatively low rate, the tracer washout was masked by the weight of the progressive radioactivity accumulation in the earlier time points. Similarly, the impact of these same features was not enough to be detectable when the whole cancer lesion was analyzed as in most studies analyzing the average kinetics of tracer accumulation in cancer [22]. This phenomenon largely reflected the combination of two different patterns in response to the delay of the regression line setup. On one side the number of releasing voxels progressively increased while preserving the washout rate. On the other side, the accumulating voxels showed a slight, yet significant, increase in the accumulation rate.

The heterogeneous tracer handling of tumor tissue documented by escalating the spatial resolution down to the voxel size closely agrees with the well documented heterogeneity of cell type populating the lesion [23,24], and with the relatively preferential localization of inflammatory cells in proximity of tumor borders [25]. Indeed, while it is universally accepted that cancer cells display an irreversible FDG accumulation [26,27], a similar consensus also applies to granulocytes and macrophages, whose tracer-releasing feature has been frequently proposed as a possible tool to differentiate cancer from inflammation [28].

## Clinical implications

A small fraction of cancer volume showed a measurable FDG washout. This finding suggests that the late tracer release should scarcely affect image quality even when acquisition is postponed beyond the 50 minutes indicated by the current guidelines [2-4]. Nevertheless, this evidence might improve the capability of PET/CT in verifying the effectiveness of treatments, such as immunotherapy, aiming to improve the host inflammatory response to cancer. Several studies report a significant increase of compartmental analysis in attributing the tracer uptake to the inflammatory infiltrates promoted by immunotherapy vs the progression of the disease [29,30]. However, the complexity and time-consuming nature of dynamic imaging so far hampered the clinical exploitation of this approach. Nevertheless, setting up voxel-resolved parametric maps delineating the trend of tumor/blood radioactivity ratio at the late time typical of tracer washout might represent a potential window to render this analysis feasible in the routine practice. Indeed, the accuracy of this approach in providing an accurate Ki value is still matter of debate [31], its capability to identify a negative slope remaining valid when the input function is not fully determined when limited to its latest time points [32]. Obviously further studies are needed to characterize the capability of this index to identify the localization of inflammatory infiltrates.

# Limitations

Several limitations of the present study should be carefully considered. On the one side, our study focused on lung cancer aiming to obtain a full and simultaneous definition of time-concentration curve in both the arterial blood and in the tumor lesion. Obviously, this localization hampers the image quality due to respiratory motion. Nevertheless, the inevitable blurring should have systematically affected all images uniformly after the steady state due to the relatively long and constant time of acquisition.

On the other side, a second limitation refers to the missing analysis of cell populations and their difference between releasing and accumulating voxels. Due to the high complexity of this evaluation, this task was not attempted, although stroma elements were obviously represented in all harvested lesions. Nevertheless, this limitation only partially hampers the nature of our proof-of-concept study aimed to identify the potential capability of voxel-resolved images to recognize heterogeneous tracer kinetics within the cancer tissue.

# Conclusions

In the present study, delaying the starting point of Patlak graphical analysis indicates a heterogeneous FDG kinetics within lung cancer lesions. This evidence can only be obtained by voxel-wise parametric maps limited to the analysis of the latest time points of dynamic acquisition. This finding indicates that expanding the analysis of FDG uptake in the time-domain might improve the informative content of PET/CT imaging adopting protocols compatible with the daily nuclear medicine practice after the introduction of novel scanners characterized by large field of view and high spatial resolution.

The potential of this time-resolved approach to FDG imaging might be of relevance for the evaluation and monitoring of therapy effectiveness, mostly in patients exposed to immunotherapy in whom differentiating the inflammatory enhancement vs the progression of disease requests, so far, pronged and serial evaluations.

14| ID | Disease site | Histology | VOI (mm3) | TV (mm3) | TV/VOI (%) | SUV (mean± std) | SUV [min, max] |
|---|---|---|---|---|---|---|---|
| 1 | left/upper lobe | adenocarcinoma (G2/G3) | 25379,7 | 6676,7 | 26,3 | 3.44 ± 0.84 | [2.39, 5.98] |
| 2 | right/upper lobe | adenocarcinoma (G2); lymph nodes involved | 156756,8 | 21834 | 13,9 | 5.04 ± 1.13 | [3.31, 8.28] |
| 3 | right medium lobe | squamous cell carcinoma | 76636,7 | 28261,8 | 36,9 | 4.97 ± 1.27 | [3.19, 7.97] |
| 4 | left/upper lobe | adenocarcinoma | 109481 | 21108,3 | 19,3 | 9.82 ± 2.32 | [6.66, 16.5] |
| 5 | right/upper lobe | adenocarcinoma (G2/G3) | 30190,2 | 3068,8 | 10,2 | 8.47 ± 2.16 | [5.6, 13.95] |
| 6 | left/bottom lobe | adenocarcinoma; lymph nodes involved | 116447,9 | 22186,5 | 19,1 | 3.42 ± 0.85 | [2.27, 5.67] |
| 7 | right upper lobe | adenocarcinoma | 180270,4 | 56938,4 | 31,6 | 18.20 ± 4.55 | [12.16, 30.41] |
| 8 | left/upper lobe | squamous cell carcinoma | 526877,2 | 184189,3 | 35 | 12.47 ± 2.45 | [7.5, 18.75] |
| 9 | left/upper lobe | squamous cell carcinoma | 323964,1 | 47172,2 | 14,6 | 13.79 ± 3.13 | [9.26, 23.13] |
| 10 | left/upper lobe | adenocarcinoma adenocarcinoma | 101891,9 | 13000,9 | 12,8 | 3.19 ± 0.75 | [2.39, 5.98] |
| 11 | right medium lobe | squamous cell carcinoma | 119433,8 | 25649,2 | 21,5 | 3.78 ± 1.00 | [2.82, 7.05] |

**Table 1**: Population description: tumor localization and final diagnosis at the histological examination of the harvested lesion; average volume, average and maximal SUV values of all detected lesions.

1724. Altorki NK., Markowitz GJ., Gao D., Port JL., et al. The lung microenvironment: an important regulator of tumour growth and metastasis. Nat Rev Cancer. 2019; https://doi.org/10.1038/s41568-018-0081-9

25. Staal-van den Brekel AJ, Thunnissen FB., Buurman WA., Wouters EF. Expression of E-selectin, intercellular adhesion molecule (ICAM)-1 and vascular cell adhesion molecule (VCAM)-1 in non-small-cell lung carcinoma. Virchows Arch. 1996; https://doi.org/10.1007/BF00192923

26. Shankar LK., Hoffman JM., Bacharach S., et al. National Cancer Institute. Consensus recommendations for the use of 18F-FDG PET as an indicator of therapeutic response in patients in National Cancer Institute Trials. *J. Nucl. Med.* 2006; 47(6): 1059:1066.

27. Sambuceti G., Cossu V., Bauckneht M., Morbelli S., et al. $^{18}$F-fluoro-2-deoxy-d-glucose (FDG) uptake. What are we looking at? Eur J Nucl Med Mol Imaging. 2021; https://doi.org/10.1007/s00259-021-05368-2

28. Schillaci O. Use of dual-point fluorodeoxyglucose imaging to enhance sensitivity and specificity. Semin Nucl Med. 2012; https://doi.org/10.1053/j.semnuclmed.2012.02.003

29. Wang D., Zhang X., Liu H., Qiu B., et al. Assessing dynamic metabolic heterogeneity in non-small cell lung cancer patients via ultra-high sensitivity total-body [$^{18}$F]FDG PET/CT imaging: quantitative analysis of [$^{18}$F]FDG uptake in primary tumors and metastatic lymph nodes. Eur J Nucl Med Mol Imaging. 2022; https://doi.org/10.21037/qims-22-725

30. Wang D., Qiu B., Liu Q., Xia L., et al. Patlak-Ki derived from ultra-high sensitivity dynamic total body [$^{18}$F]FDG PET/CT correlates with the response to induction immuno-chemotherapy in locally advanced non-small cell lung cancer patients. Eur J Nucl Med Mol Imaging. 2023; https://doi.org/10.1007/s00259-023-06298-x

31. Chen Z., Cheng Z., Duan Y., Zhang Q., et al. Accurate total-body $K_i$ parametric imaging with shortened dynamic $^{18}$F-FDG PET scan durations via effective data processing. Med Phys. 2023; https://doi.org/10.1002/mp.15893

32. Zuo Y., Qi J., Wang G. Relative Patlak plot for dynamic PET parametric imaging without the need for early-time input function. *Phys. Med. Biol.* 2010; https://doi.org/10.1088/1361-6560/aad444

# Statements & Declarations

## Funding

The activity of MP and CC has been performed within the framework of the Next Generation EU project "RAISE - Robotics and AI for Socio-economic Empowerment". This research was funded by the Italian Ministry of Health 5 × 1000 2018–2019 with the project "Hybrid Hub (H2UB): Modelli cellulari e computazionali, micro e nanotecnologie per la personalizzazione di terapie innovative", T4-AN-10; by the Italian Ministry of Health 5 × 1000 2020 with the project "IPO: imaging parametrico in oncologia", and by the program Ricerca Corrente 2023, line "Guest-Cancer Interactions".

## Competing Interests

The authors declare no competing interest.

## Author Contributions

DP and ED performed the analysis and wrote the paper; GB performed the analysis; FD, MB and CM collected the data and conceived the analysis; SG and CC contributed the analysis tools, performed the analysis and wrote the paper; MP and GS conceived and designed the analysis and wrote the paper.

## Data Availability

The datasets generated during and/or analyzed during the current study are available from the corresponding author on reasonable request.

# Ethics approval

The patients were included in the context of a clinical trial approved by the local ethical committee and registered as NCT02475382.

# Consent to participate

Informed consent was obtained from all individual participants included in the study.

# Consent to publish

All patients provided the signed informed consent form for participation and publication of anonymized images.